\documentclass[multphys,vecphys]{svmult}

\usepackage{makeidx}         % allows index generation
\usepackage{graphicx}        % standard LaTeX graphics tool
\usepackage{multicol}        % used for the two-column index
\usepackage[bottom]{footmisc}% places footnotes at page bottom
\usepackage{natbib}
\makeindex             % used for the subject index
\begin{document}

\title*{Scalable N-body code for the modelling of early-type galaxies}

\author{Emilie~Jourdeuil and Eric~Emsellem}

\institute{Centre de Recherche Astrophysique de Lyon (CRAL)\\
   Universit\'e Lyon~I, Observatoire de Lyon \\
 9 avenue Charles Andr\'e\\
 69561 Saint Genis Laval Cedex\\
\texttt{emilie@obs.univ-lyon1.fr, emsellem@obs.univ-lyon1.fr}}

\maketitle
\begin{abstract}
Early-type galaxies exhibit a wealth of photometric and
dynamical structures. These signatures are fossil records of their
formation and evolution processes. 
In order to examine these structures in detail, we build models aimed
at reproducing the observed photometry and kinematics. 
The developed method is a generalization of
the one introduced by \cite{sye96}, consisting in an N-body representation,
in which the weights of the particles are changing with time. Our
code is adapted for integral-field spectroscopic data, and is able to reproduce
the photometric as well as stellar kinematic data of observed galaxies. We apply
this technique on SAURON data of early-type galaxies, and present
preliminary results on NGC~3377.
\end{abstract}

\section{Introduction}
{\bf Goal}
Elliptical galaxies are presumed to be the product of galaxy merging, where different mass ratios
result in different classes of elliptical galaxies (disky, boxy\dots) \cite{naa99}. This
seems consistent with the hierarchical scenario of galaxy formation
\cite{pre74}, although it is clear that more standard collapse processes and secular 
evolution contribute to the shaping of galaxies.
There are many different and often complementary approaches to constrain these scenarios. 
The one we adopted is to study the structures of nearby elliptical galaxies in detail, 
in order to probe the signatures of their evolution and to be able to trace their formation
history. This fossil search requires data both on the chemical and
dynamical status of the stellar components, and if present of the gaseous
system, both being provided by spectroscopy. \\

\hspace{-.5cm}{\bf 3D spectroscopy} Spectroscopy can indeed deliver information on the stars'
dynamics and chemistry via the analysis of their absorption lines. Two-dimensional spatial
coverage provided by integral-field spectroscopy is critical to homogeneously sample
an extended target, while more traditional long-slit spectroscopy generally restricts our view 
to a few a priori fixed axes. 
A qualitative assessment of the obtained maps is clearly not sufficient to fully
address the issues mentioned above. We need to go further by
modelling the observed galaxies using constraints retrieved from 3D spectroscopy, such as
luminosity, stellar velocity distribution and stellar populations. As emphasized by  \cite{bin05},
3D spectroscopy seems to be essential to properly constrain the dynamics of
the galaxy under scrutiny.\\

\hspace{-.5cm}{\bf Existing models}
One of the key goal of galaxy modelling is to retrieve the full distribution function (DF), 
{\it i.e.} the density of stars in phase-space.
There are already several known methods to address this problem,
each of them having both advantages and drawbacks. We can distinguish different
classes of techniques~:
\begin{itemize}
\item DF-based method, where models are often restricted to simple geometries.
\item Moment-based method, which consists in solving Boltzmann and
Poisson equations via the use of a closed system of relations (Jeans equations). The main issue here is that the final DF may not be positive everywhere.
\item Orbit-based method, or Schwarzschild modelling \cite{sch79}. Libraries
 of orbits are built within a fixed potential, and each orbit is weighted
 as to reproduce the observed galaxy (photometry and kinematics).
\item Particle-based method, where the particles represent groups of stars.
 One has to guess the right initial conditions which will evolve in a configuration
 resembling the chosen galaxy.
\end{itemize}
\section{Method}
The method we wish to present here is an hybrid scheme between the Schwarzschild and N-body techniques.
It has been first proposed and developed by \cite{sye96} and consists in an N-body realization where
the weight of each particle is gradually changed in order to fit the observables.\\

\hspace{-.5cm}{\bf Previous applications}
This method has been tested by \cite{sye96} on fake galaxy models, adjusting
the photometry alone. It has been recently applied on the Milky Way by \cite{bis04}, 
fitting the photometry of a frozen snapshot extracted from a full N-body
simulation\footnote{An extended independent implementation of this algorithm has been recently presented by de Lorenzi et al. 2007: see their paper for details}.\\

\begin{figure}
\centering
\includegraphics[width=0.6\textwidth]{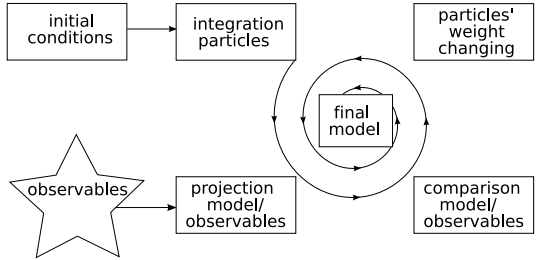}
\caption{Algorithm of the Syer \& Tremaine method.}
\label{jou:fig1}
\end{figure}

\hspace{-.5cm}{\bf Algorithm}
The algorithm we use is illustrated in Fig.~\ref{jou:fig1}.
Particles start from (N-body) initial conditions and are integrated along their orbits. 
Observables are used as reference input to the modelling, particles being projected 
as to mimic these observables. A weight prescription is derived from the
difference between the model and the data, 
this prescription being then applied on each particle accordingly. 
This procedure (integration, projection, comparison, weight changing) is
repeated until the weights converge.
\\

\hspace{-.5cm}{\bf Initial conditions}
The initial conditions have to properly sample the galaxy, {\it ie} in position/velocity space or using integrals of motions. This is not an easy task when we do not {\it a priori} know the internal structure of the galaxy we wish to model.
If we want to model real galaxies, we have to find
appropriate initial conditions, without a priori knowing its DF. We thus first
build an initial mass distribution, by using the known photometry and assuming a
constant mass-to-light ratio. This is achieved here via the Multi Gaussian Expansion (MGE) 
proposed by \cite{mon92}, which allows us to decompose the mass distribution 
into a sum of tri-dimensional Gaussians, and to derive the corresponding gravitational
potential analytically \cite{ems94}. Jeans equations are then used to obtain a first guess of
the dynamics and set up initial conditions.\\

\hspace{-.5cm}{\bf Integration}
In a first version of the code, we keep the potential steady. The code has been
developed in a very modular and flexible fashion, such as to allow the easy
implementation of a self-consistency module~: this will be described in 
a forthcoming paper. As we need to evaluate the observables of all particles at the same time, 
we chose a second-order synchronized leapfrog scheme, where positions and velocities can be evaluated
simultaneously. The leapfrog technique is perhaps not the best choice in terms of accuracy, but 
was found sufficient for this first development step. Other schemes (e.g. Runge-Kutta) can be easily
implemented.
Adaptive time-steps have been included, in order to optimally sample all 
the orbits. An independent integration of each particle would result in a very
inefficient algorithm. We thus chose to follow the algorithm described in 
the N-body code GADGET \cite{spr01}, in which particles advance in bunches 
so that they are always distributed in a tight time range around the current time.\\

\hspace{-.5cm}{\bf Observables}
Our code has been designed around two new main items.
First, we do not restrict ourselves to fit the photometry~: the code also allows the fitting of the kinematics, via the use of Line of Sight Velocity
Distributions (LOSVDs). Second, the code has been developed as to permit the use of 
3D spectroscopic data, including spatial adaptive binning (Voronoi bins, see \cite{cap03}). We have thus used the data obtained with the SAURON spectrograph (WHT,
Canary Islands) for its survey of early-type galaxies \cite{dez02,ems04}.\\

\hspace{-.5cm}{\bf Prescription}
The heart of the code is the prescription developed by \cite{sye96}, which
rules the weight changing of each particle so that the cumulated
observables reproduce the observations~:
\begin{equation}
  \frac{dw_i(t)}{dt} = -\epsilon w_i(t)\sum_{j=1}^J\frac{K_{ij}}{Z_j}\Delta_j
\label{jou:eq1}
\end{equation}
where $i$ describes the particles, $j$ the observables, $w$ the weights,
$K_{ij}$ the contribution of the particle $i$ to the observable $j$, and $\Delta_j
\equiv y_j(t)/Y_j -1$ the contribution of all the particles to this same observable. $\epsilon$ is the strength of the weight changing, $w_i$
avoids the weights to be negative when $w_i$ approaches $0$. 
The input parameters are the following~: 
\begin{itemize} \item $N$ is the number of particles,
\item $d_{time}$ is the time during which the observables are averaged,
\item $\epsilon$ is the strength parameter mentioned in Eq.~\ref{jou:eq1},
\item $\alpha$ is the smoothing parameter of the observables, used in the equation~:
\begin{equation}
      \tilde\Delta_j(t) = \alpha \int_0^\infty\Delta_j(t-\tau)e^{-\alpha \tau}d\tau
    \end{equation}
so that $\tilde\Delta_j(t)$ replaces $\Delta_j(t)$ in Eq.~\ref{jou:eq1}.
\end{itemize}
The choice of these parameters is crucial, each galaxy requiring an
appropriate set.
\section{Results}

\begin{figure}
\centering
\includegraphics[width=0.7\textwidth]{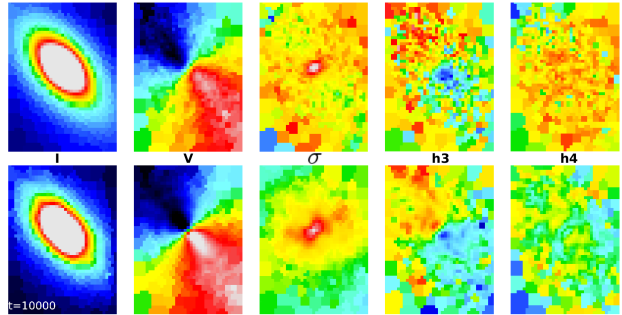}
\caption{$I, V, \sigma, h_3, h_4$ from the observations (top) and the model (bottom) of NGC~3377}
\label{jou:fig2}
\end{figure}

A preliminary model has been obtained on the nearby disky galaxy
NGC~3377. In Fig.~\ref{jou:fig2} we illustrate the fit of the photometry and LOSVDs 
obtained with the SAURON instrument on this galaxy, for a given mass
(MGE) distribution. The maps are globally well reproduced ($I$, $\sigma$, $h_3$).
Some remaining discrepancy in the velocity field is probably due to an observed asymmetry in
the kinematics of this galaxy (tilt between the photometric and kinematic
minor-axis). The relatively high $h_4$ observed level is not reached by the
model~: this may be partly related to a template mismatch effect which
mostly affects the even moments of the LOSVD.

\section{Conclusion and perspectives}
Although the results shown here are very preliminary,
they illustrate the possibility of fitting the kinematics,
which is an important step in the development of such a code. 
It has also been designed to adjust complex data such as the
one provided by 3D spectrographs. A few improvements have been or are now being implemented~:
\begin{itemize}
\item Self-consistency~: it is technically easy to replace the current integration 
module with a self-consistent scheme. The full implications of a continuous weight changing
remains however unclear and have to be examined.
\item Addition of a central black hole~: this would allow the use of higher
spatial resolution data of nearby galaxies.
\item Stellar populations~: this new class of observables would allow us
to link the dynamics and chemistry of the stellar component.
\end{itemize}
%
%
% BibTeX users please use
 \bibliographystyle{abbrv}
 \bibliography{jourdeuil}

\begin{thebibliography}{10}

\bibitem{bin05}
J.~{Binney}.
\newblock {\em {\it{MNRAS}}}, 363:937--942, Nov. 2005.

\bibitem{bis04}
N.~{Bissantz}, V.~P. {Debattista}, and O.~{Gerhard}.
\newblock {\em {\it{ApJ}}}, 601:L155--L158, Feb. 2004.

\bibitem{cap03}
M.~{Cappellari} and Y.~{Copin}.
\newblock {\em {\it{MNRAS}}}, 342:345--354, June 2003.

\bibitem{dez02}
P.~T. {de Zeeuw}, M.~{Bureau}, E.~{Emsellem}, R.~{Bacon}, C.~{Marcella
  Carollo}, Y.~{Copin}, R.~L. {Davies}, H.~{Kuntschner}, B.~W. {Miller},
  G.~{Monnet}, R.~F. {Peletier}, and E.~K. {Verolme}.
\newblock {\em {\it{MNRAS}}}, 329:513--530, Jan. 2002.

\bibitem{ems04}
E.~{Emsellem}, M.~{Cappellari}, R.~F. {Peletier}, R.~M. {McDermid}, R.~{Bacon},
  M.~{Bureau}, Y.~{Copin}, R.~L. {Davies}, D.~{Krajnovi{\'c}}, H.~{Kuntschner},
  B.~W. {Miller}, and P.~T. {de Zeeuw}.
\newblock {\em {\it{MNRAS}}}, 352:721--743, Aug. 2004.

\bibitem{ems94}
E.~{Emsellem}, G.~{Monnet}, and R.~{Bacon}.
\newblock {\em {\it{A\&A}}}, 285:723--738, May 1994.

\bibitem{mon92}
G.~{Monnet}, R.~{Bacon}, and E.~{Emsellem}.
\newblock {\em {\it{A\&A}}}, 253:366--373, Jan. 1992.

\bibitem{naa99}
T.~{Naab}, A.~{Burkert}, and L.~{Hernquist}.
\newblock {On the Formation of Boxy and Disky Elliptical Galaxies}.
\newblock {\em {\it{ApJ}}}, 523:L133--L136, Oct. 1999.

\bibitem{pre74}
W.~H. {Press} and P.~{Schechter}.
\newblock {\em {\it{ApJ}}}, 187:425--438, Feb. 1974.

\bibitem{sch79}
M.~{Schwarzschild}.
\newblock {\em {\it{ApJ}}}, 232:236--247, Aug. 1979.

\bibitem{spr01}
V.~{Springel}, N.~{Yoshida}, and S.~D.~M. {White}.
\newblock {\em \it {New Astronomy}}, 6:79--117, Apr. 2001.

\bibitem{sye96}
D.~{Syer} and S.~{Tremaine}.
\newblock {\em {\it{MNRAS}}}, 282:223--233, Sept. 1996.

\end{thebibliography}
%%%%%%%%%%%%%%%%%%%%%%%%%%%%%%%%%%%%%%%%%%%%%%%%%%%%%%%%%%%%%%%%%%%%%%  }

%%%%%%%%%%%%%%%%%%%%%%%%%%%%%%%%%%%%%%%%%%%%%%%%%%%%%%%%%%%%%%%%%%%%%%

\printindex
\end{document}